\documentclass[12pt,epsfig]{article}
\usepackage{amsmath}
\usepackage{graphicx}
\usepackage{epstopdf}
\usepackage{color}
\usepackage{titlesec}
\usepackage{inputenc}
\usepackage{cases}
\titleformat*{\section}{\large\bfseries\sffamily}
\titleformat*{\subsection}{\small\bfseries\sffamily}
\begin{document}	
	\begin{center}
		\textbf {Clustering of eclipsing binary light curves through functional principal component analysis}\\
		\vskip.2in Soumita Modak$^{1,*}$, Tanuka Chattopadhyay$^{2}$\\
		{\small and}\\
		Asis Kumar Chattopadhyay$^{3}$\\
		\vskip.1in
		$^{1}$Department of Statistics, Basanti Devi College\\ affiliated with  University of Calcutta\\
		147B, Rash Behari Ave, Kolkata- 700029, India\\
		$^{*}$email: soumitamodak2013@gmail.com\\
		Orcid id: 0000-0002-4919-143X\\
		\vskip.1in
		$^{2}$Department of Applied Mathematics, University of Calcutta\\
		92 A.P.C. Road, Kolkata- 700009, India\\
		email: tanuka@iucaa.ernet.in\\
		\vskip.1in
		$^{3}$Department of Statistics, University of Calcutta\\35 Ballygunge Circular
		Road, Kolkata- 700019, India\\
		email: akcstat@caluniv.ac.in\\
	\end{center}
	
	\begin{abstract}
		In this paper, we revisit the problem of clustering 1318 new variable stars found in the Milky way. Our recent work distinguishes these stars based on their light curves which are univariate series of brightness from the stars observed at discrete time points. This work proposes a new approach to look at these discrete series as continuous curves over time by transforming them into functional data. Then, functional principal component analysis is performed using these functional light curves. Clustering based on the significant functional principal components reveals two distinct groups of eclipsing binaries with consistency and superiority compared to our previous results. This method is established as a new powerful light curve-based classifier, where implementation of a simple clustering algorithm is effective enough to uncover the true clusters based merely on the first few relevant functional principal components. Simultaneously we discard the noise from the data study involving the higher order functional principal components. Thus the suggested method is very useful for clustering big light curve data sets which is also verified by our simulation study.\\
		
		
	\end{abstract}
	
	\section{Introduction}	\label{intro}
	An eclipsing binary (E) is a special kind of variable star consisting of a binary star system with each star eclipsing the other periodically, which makes variation in their total brightness of light over time. Classification of Es using their varying light curves (LCs) reveals significant information regarding stellar structure and
	evolution (Malkov et al. 2007; Kirk
	et al. 2016; Kochoska et al. 2017; Mowlavi
	et al. 2017; S\"{u}veges et al. 2017; Modak 2019; Modak et al. 2020; Modak 2022).
	The data set under our study contains observations on 1318 new variable stars (Miller et al. 2010) covering 0.25 square degree region of the Galactic plane centered on Galactic coordinates (latitude, longitude) of $(330.94$, $-2.28)$
	degree. Subjective study (Miller et al. 2010) based on the appearance of observed LCs of the stars hinted at four possible groups, viz. Algol type (EA), Beta Lyrae (EB), W Ursae Majoris (EW) and un-categorized pulsating stars (PUL). 
	However, automated or quantitative methods are preferable to reduce such subjectivity by the researchers which caused uncertainty and ambiguity in the classification of many stars. Therefore, in an attempt to improve their results and to crosscheck the existing groups of stars we propose, in Modak et al. (2020), the $k$-medoids clustering (Kaufman and Rousseeuw 2005) with the complexity invariance distance (CID) (Prati and Batista 2012; Batista et al. 2014; Wei 2014). It 
	 resulted in two physically interpretable groups of Es. To check the robustness of the resulting groups, this paper
	revisits the problem of clustering the stars based on their LCs from a new perspective.
	
	Established literature give an overview of analyzing astronomical data sets in the classical multivariate set-up, be it time series or cross-sectional data (Matijevi\v{c} et al. 2012; Chattopadhyay et al. 2016; Kirk et al. 2016;  Kochoska et al. 2017; Modak et al. 2017; Mowlavi et al. 2017; S\"{u}veges et al. 2017; Bandyopadhyay and Modak 2018; Modak et al. 2018, 2020; Modak 2019; Modak and Bandyopadhyay 2019; Modak 2021a, 2021b, 2022). In contrast, the novelty of our present work lies in dealing with time series not as a multivariate data set, where the variables are measured at discrete points ordered in time, but rather as continuous curves of time called functional data. Various methods are available for clustering of functional data sets (for example, see, Jacques and Preda 2014; Delaigle et al. 2019). Here LCs, which are time series observed at discrete time points on the brightness of 1318 stars, are transformed into functional data on which we perform the functional principal component analysis (FPCA, see Ramsay and Silverman 2002, 2005; Ramsay et al. 2009 and references therein). Hierarchical clustering (Ward 1963; Kaufman and Rousseeuw 2005) based on the significant functional principal components (FPCs), extracted through FPCA, reveals two inherent groups of Es. One group indicates bright and massive systems of Es while the other corresponds to fainter and less massive systems.
	
	It exhibits consistency with the clustering results found by Modak et al. (2020). Our present method is shown to be superior in terms of clustering accuracy measure the `connectivity' (Handl et al. 2005) and the well-separated average LCs obtained from the resulting groups. The proposed approach also solves the concerns of big LC data sets. Because FPCA is used as a strong dimension reduction technique, wherein only the first few significant FPCs can sufficiently extract the relevant information available in the original data. Simultaneously we can reduce noise by discarding the higher order irrelevant FPCs. Clustering based on the important FPCs leads to the desired results. These facts are confirmed through our simulation study, which establishes our proposed method as an effective LC-based classifier.
	
	The paper is organized as follows. Section 2 discusses the construction of functional data from the discrete observations and Section 3 gives the theory of FPCA. Our data and clustering method with simulation study are explained respectively in Sections 4 and 5. Section 6 shows our findings and explains the physical interpretations of the output, whereas Section 7 concludes. 
	\section{Functional data}
	For a data set of size $N$ with the $i$-th ($i=1,\ldots,N$) observation having a set of $n$ discrete measured values $\mathbf{y}_i=(y_{i1},\ldots,y_{in})'$ (vectors are indicated by bold symbol throughout the paper), we transform these discrete values into a function $x_i$ with a continuous curve $x_i(t)$ corresponding to the $i$-th observation in the sense that the value of  $x_i(t)$ is computable for any given value of $t$ (in our study, $t$ is the time variable). Thus $\{x_i(\mathbf{t})\}_{i=1}^{N}$ represent the functional data for the given data set, where $x_i(\mathbf{t})$ is the functional data vector corresponding to the observed data vector $\mathbf{y}_i$ at the vector $\mathbf{t}=(t_1,\ldots,t_n)'$ of argument values of $t$ (Ramsay and Silverman 2002, 2005; Ramsay et al. 2009). So basically, we try to fit the model, with the observation matrix
	\begin{equation*}
		Y=(y_{ij})=(\mathbf{y}_1,\ldots,\mathbf{y}_N)',
	\end{equation*}
	the corresponding functional data matrix 
	\begin{equation*}
		X=(x_{i}(t_j))=(x_1(\mathbf{t}),\ldots,x_N(\mathbf{t}))'
	\end{equation*}
	and the unobserved random error matrix $\epsilon =(\epsilon_{ij})=(\boldsymbol{\epsilon}_1,\ldots,\boldsymbol{\epsilon}_N)'$, as follows:\\
	\begin{equation}
		Y=X+\epsilon.
	\end{equation}
	\subsection{Estimation of function} 
	Estimation of the function $x_i$ is done in the same way for each $i$, hence the subscript $i$ is suppressed to avoid the notational clutter, i.e. from now on $
	\mathbf{y}_i, x_i(\mathbf{t})$ and $\boldsymbol{\epsilon}_i$ are respectively denoted by $\mathbf{y}, x(\mathbf{t})$ and $\boldsymbol{\epsilon}$. Therefore,
	\begin{equation}\label{vectorobn}
		\mathbf{y}=x(\mathbf{t})+\boldsymbol{\epsilon}
	\end{equation}
	with E$(\mathbf{y})=x(\mathbf{t})$, since it is assumed that $\text{E}(\boldsymbol{\epsilon})=\mathbf{0}$ and $x$ is treated as a fixed effect function, and Disp$(\mathbf{y})=\text{Disp}(\boldsymbol{\epsilon})=\Sigma$ (here `E' and `Disp'  denote respectively the expectation and the dispersion).
	
	Given $\mathbf{y}$ and $\lambda$, the function $x$ is estimated by minimizing (Ramsay and Silverman 2005)
	\begin{equation}\label{criteria1}
		[\mathbf{y}-x(\mathbf{t})]'W[\mathbf{y}-x(\mathbf{t})]+\lambda\int \bigg[\frac{d^2}{dt^2}x(t)\bigg]^2dt,
	\end{equation}
	provided $\int[\frac{d^2}{dt^2}x(t)]^2dt$ exists and $W=\Sigma^{-1}$. Equation~\eqref{criteria1} considers a trade-off between smoothness and data fit wherein $\lambda$ $(>0)$ is a smoothing parameter. $\lambda=0$ reduces Equation~\eqref{criteria1} to the weighted least square fitting, which for $\Sigma=\sigma^2I_n$ further boils down to the least square fitting.
	\subsection{Representing function by basis functions} 
	The function $x$ is approximated by a
	linear combination of $K$ (say) known basis functions $\phi_{1},\ldots,\phi_{K}$ such that $x(t_j)=\sum_{\tilde{k}=1}^{K}c_{\tilde{k}} \phi_{\tilde{k}}(t_j)=\boldsymbol{\phi}'(t_j)\mathbf{c}$, with $\boldsymbol{\phi}(t_j)=(\phi_1(t_j),\ldots,\phi_{K}(t_j))'$ and $\mathbf{c} = (c_{1},\ldots,c_{K})'$, for all $j=1,\ldots,n$. Thus, Equation~\eqref{vectorobn} can be written in the following form 
	\begin{equation}
		\mathbf{y}= \Phi \times \mathbf{c} + \boldsymbol{\epsilon}, 
	\end{equation}
	where $\mathbf{c}$ 
	is a vector of unknown constants to be estimated and $\Phi = (\boldsymbol{\phi}(t_1),\ldots,\boldsymbol{\phi}(t_n))'$
	is the matrix of the known basis functions. The choice of $K$ is provided by the analyst.
	
	Thus, the problem of estimating the function $x$ by minimizing Equation \eqref{criteria1} boils down to estimation of the unknown coefficient vector $\textbf{c}$ by minimizing 
	\begin{equation}\label{criteria2}
		(\mathbf{y}-\Phi\mathbf{c})'W(\mathbf{y}-\Phi\mathbf{c})+\lambda\mathbf{c}'R^*\mathbf{c},
	\end{equation}
	where $\mathbf{c}'R^*\mathbf{c}=\int [\frac{d^2}{dt^2}x(t)]^2dt$ and $R^*=\int \frac{d^2}{dt^2}\boldsymbol{\phi}(t) \frac{d^2}{dt^2}\boldsymbol{\phi}'(t)dt$ which is called the penalty matrix. Therefore, estimated $x$ is obtained as 
	\begin{equation}
		\hat{x}(\mathbf{t})=\Phi \times \hat{\mathbf{c}}
		=\Phi[\Phi'W\Phi+\lambda R^*]^{-1}\Phi'W\mathbf{y}.
	\end{equation}
	In our analysis, an order four B-spline (de Boor 2001) basis function expansion with knots at the sampling
	points is chosen, i.e. $\phi_{\tilde{k}}$ ($\tilde{k}=1,\ldots,K$) is selected to be a polynomial of degree three, which minimizes Equation \eqref{criteria2} with respect to $\mathbf{c}$. The value of $K=n+p-1$, where $n$ is the number of sampling points and $p$ is the degree of the chosen polynomial, here $K=274$ (see, Step (\textbf{S1}) under Section \ref{4.1}). The
	penalty matrix is estimated by Cholesky decomposition and numerical quadrature approximation (Stoer and Bulirsch 2002; Ramsay and Silverman 2005). We assume $W=I_n$ and the value of $\lambda$  is evaluated by the measure of generalized cross-validation (Craven and Wahba
	1979; Gu 2002).
	\section{FPCA}
	In classical multivarite set-up, for an $n-$variate random observation vector $\mathbf{y}$ with Disp$(\mathbf{y})=\Sigma$, we solve the following equation:
	\begin{equation}
		\Sigma\boldsymbol{\xi}=\rho\boldsymbol{\xi},
	\end{equation}
	for nonnegative eigenvalues $\rho$ and nonzero eigenvectors $\boldsymbol{\xi}$. Let $\rho_1\geq\ldots\geq\rho_n\geq0$ be the eigenvalues of $\Sigma$ with $\rho_l$ being the last nonzero eigenvalue and $\boldsymbol{\xi}_1,\ldots,\boldsymbol{\xi}_n$ be the corresponding orthonormal set of eigenvectors. Then the $m-$th principal component of $\mathbf{y}$ is obtained as
	\begin{equation}
		f_m=\sum_{j=1}^{n}\xi_{mj}y_j=\boldsymbol{\xi}_m'\mathbf{y},\hspace{.1in} m=1,\ldots,l.
	\end{equation}  
	
	Now, the functional version of the above principal component analysis is studied based on the functional data vector $x(\mathbf{t)}$ corresponding to $\mathbf{y}$ with covariance function  $v(s,t)=\text{cov}(x(s),x(t))$. Here we solve an analogous eigen equation
	\begin{equation}
		V{\xi}(s)=\rho{\xi}(s),
	\end{equation}
	for nonnegative eigenvalues $\rho$ and nonzero eigenfunctions ${\xi}(t)$, where $V$ is an integral transform of $\xi$ called `covariance operator' with $V\xi(s)=\int v(s,t)\xi(t)dt$. As the function $x$ can be defined for any value of $t$, in contrast to the multivariate PCA, FPCA can have infinite number of pairs of eigenvalue-eigenfunction. Let $\rho_1\geq\rho_2\geq\ldots\geq0$ be the eigenvalues of the operator $V$ with $\rho_l$ being the last nonzero eigenvalue and $\xi_1(s),\xi_2(s),\ldots$ be the corresponding orthonormal set of eigenfunctions. Then the $m-$th functional principal component (FPC) of $\mathbf{y}$ is obtained as
	\begin{equation}
		f_m=\int \xi_m(s) x(s) ds=\int \xi x,\hspace{.1in} m=1,\ldots,l.
	\end{equation}  
	\section{Data}
	The observed data set (Miller et al. 2010) has LCs (relative flux variation in R-band over time measured in Heliocentric Julian Date abbreviated to HJD) of 1318 variable stars in our Galaxy along with R-band magnitude (R), colors (B-R, R-I) and period (P). Other computed variables are I (R minus R-I), B-I (B-R plus R-I) and B (B-I plus I). For all the stars, the observed LCs are unevenly spaced of different lengths (ranging from 130 to 264, except one LC having length 5) with values at different time points and the period is varying from several hours to several weeks.\\
	\subsection{Data processing}
	\label{4.1}
	Data on which clustering is performed undergo the following process:\\
	(\textbf{S1}) Observed LCs are transformed into a full cycle over phase interval [0,1] (Percy 2007; Deb and Singh 2009; Soszy\'{n}ski et al. 2016), having observations at 272 evenly spaced phase points, as explained below briefly (for details, see, Modak et al. 2020). \\
	(i) The phased LCs are obtained by transforming the given time points into standard phases (always lie between 0 and 1) with the help of the following equation (Percy 2007; Deb and Singh 2009)
	\begin{equation}\label{standardphase}
		\text{decimal}\hspace{1 mm}\text{portion}\hspace{1 mm}\text{of}\hspace{1 mm}[(t-t_{0})/\text{P}],
	\end{equation}
	wherein $t$: time of having measurement on brightness of the star in HJD,\\
	$t_{0}$: time of the first observed maximum brightness and\\
	P: period of the star in days.\\ 
	(ii) Step (i) gives the $i^{th}$ LC over phase interval $[0,p_{i}]$, with $p_{i}$ as the maximum of standard phases for the $i^{th}$ LC, which is extended to phase interval $[0,p_i+1]\in[0,2)$ by adding $+1$ to the standard phases obtained from step (i), for all $i = 1,\ldots,1318$.\\
	(iii) We fit the linear spline (Press et al. 1992; Cassisi et al. 2012) to the LCs from step (ii) at $l'$ evenly spaced phases over [0,1] using the following formulas:\\
 (a) For star with the $i^{th}$ LC, we have $l_i$ values of the brightness function $y$ against different values of phase point $p'$ as $y_{j}=y(p'_{j})$, $j =1,\ldots,l_i$ with $p'_{j}<p'_{j+1}$ for $ j = 1,\ldots,l_i-1$.\\
 (b) The interpolating function joins the following $(l_i-1)$ linear functions 
 \begin{equation*}
 	g_{j}(p')=a_{j}y{_j}+b_{j}y_{j+1},\hspace{1 mm}\text{for}\hspace{1 mm}p'\in[p'_{j}, p'_{j+1}],\hspace{1 mm}j=1,\ldots,l_i-1.
 \end{equation*}  
 (c) $a_{j}$ and $b_{j}$ are constants which satisfy the followings:\\
  (c1.1) $g_{j}(p'_{j})=y_{j}$ and (c1.2) $g_{j}(p'_{j+1})=y_{j+1}$ for $j=1,\ldots,l_i-1$, i.e.\\
  \begin{equation*}
  a_{j}=\frac{p'_{j+1}-p'}{p'_{j+1}-p'_{j}}
 \end{equation*} 
   and 
   \begin{equation*}
   b_{j}=1-a_{j}=\frac{p'-p'_{j}}{p'_{j+1}-p'_{j}}.
  \end{equation*} 
	(iv) Finally, the lengths of all LCs are fixed at $l'=272$, after a trade-off between  extraction of relevant information from the LCs and the interpolation error in approximating the LCs, which produced the best possible clustering results (Modak et al. 2020).\\
	(\textbf{S2}) Each LC from step (\textbf{S1}) is transformed into functional data (details discussed in Section 2).\\ 
	(\textbf{S3}) We perform FPCA (see, Section 3) on the functional data set obtained from step (\textbf{S2}) and extract the FPCs to which the clustering method is applied.
	\section{Clustering}
	Hierarchical clustering, an agglomerative method (Kaufman and Rousseeuw 2005) based on the Ward's algorithm (Ward 1963), is applied to the first few significant FPCs wherein the number of clusters is obtained by the connectivity (Handl et al. 2005), which measures the tightness within the clusters using a preassigned number of nearest neighbors. This clustering accuracy measure
	takes a nonnegative value with a minimum value indicating the best possible clustering. Throughout the cluster analysis, the Euclidean norm is used to compute distances among the FPCs and to calculate the connectivity 10 nearest FPCs are considered.
	\subsection{Clustering accuracy: connectivity}
	Suppose our implemented clustering algorithm has clustered the data set consisting of $N$ data members into $k$ $(>1)$ mutually exclusive and exhausted clusters $C_1$,\ldots,$C_k$ of sizes $n_1,\ldots,n_k$ respectively, with $\sum_{c=1}^{k}n_c=N$. Let $M_{c,m}$ represent the $m-$th member of the $c-$th cluster $C_c$ or the corresponding observation in the data set for $m=1,\ldots,n_c$ and $c=1,\ldots,k$, and the distance between any two members $M$ and $M'$ of the data is given by $d(M,M')$, the Euclidean norm in our study.
	
	Now, for the clustered structure, we assign a quantity $I_{c,m}(j)$ to the member $M_{c,m}$ such that 
		\begin{numcases}
		{I_{c,m}(j) =}
		0       & if the $j-$th nearest member of $M_{c,m}\in C_c$, \nonumber\\
		1/j  & otherwise,
	\end{numcases}
	where proximity between members is measured in terms of the distance between them. The required clustering accuracy measure `connectivity' is defined as (Handl et al. 2005)
	\begin{equation}
		\text{Conn}=\sum\limits_{c=1}^{k}\sum\limits_{m=1}^{n_c}\sum\limits_{j=1}^{J}I_{c,m}(j), 
	\end{equation}
	where the value of the parameter $J$ is provided by the analyst, with $J=10$ in this paper. Clearly, Conn$\in[0,\infty)$ and as it indicates the compactness of the formed clusters, a low value is desired. Smaller its value means better is the resulting clustering.
   \subsection{Evaluation of the number of clusters}
   We perform unsupervised classification applying a hierarchical clustering algorithm, where the true value of the number of clusters ($k$) present in the data set is unknown. Therefore for different possible values of $k$, here $k=2,3,\ldots,6$, we perform the clustering and subsequently compute the value of the cluster validity index `connectivity' using the clustered data set (see, Equation 13). Finally, the value of $k$ is chosen for which the connectivity attains its minimum (discussed in Section 5.1).
	\subsection{Simulation study}
\begin{figure}
		\includegraphics[width=1\textwidth]{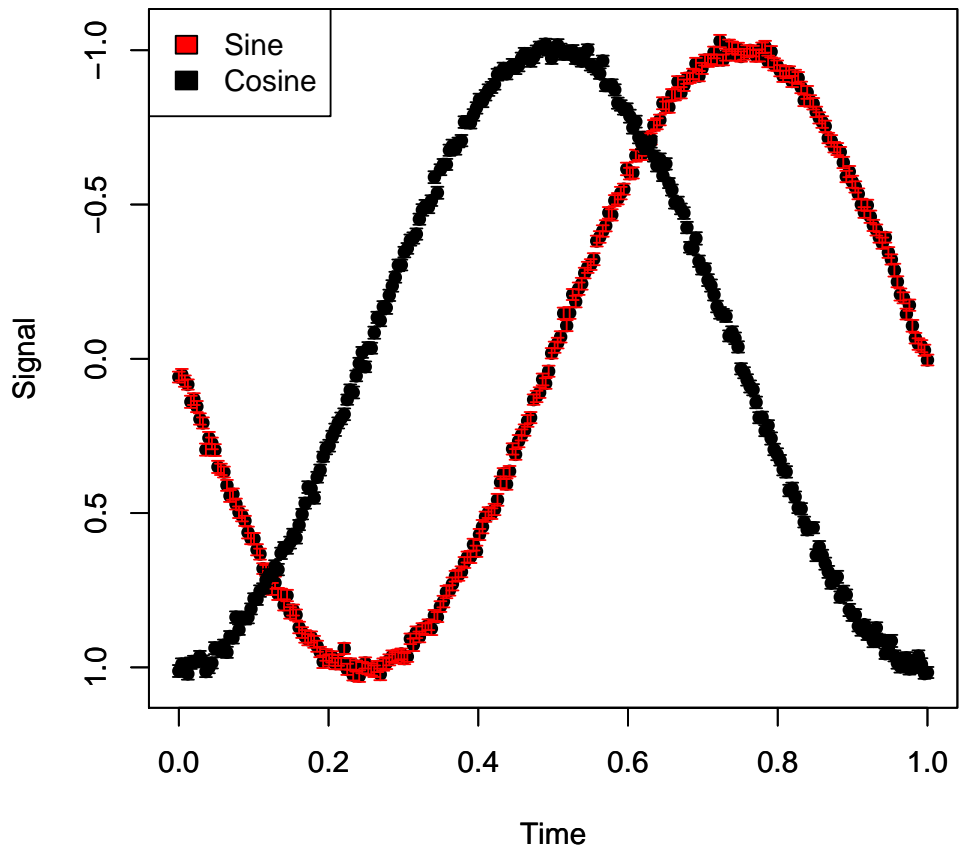}\\
		\caption{Simulated average light curves (with standard errors) of 500 members from each of the two groups generated using sine and cosine signals with
			added noise and outliers}\label{simulation1}
	\end{figure}

	We check the performance of our approach through simulated LCs which are generated using a periodic signal contaminated with noise and outliers at 250 evenly spaced time points on [0,1] (see, for details, Thieler et al. 2013, 2016 and references therein). 
	In the first case, we consider complete cycles of two different signals sine and cosine contaminated with signal-to-noise ratio (SNR) $=3$ (90\% of the noise is related to the measurement accuracies and 10\% is white noise) and $10\%$ outliers added to the measurement accuracies. It simulates 500 LCs from each of two groups both having amplitude 1 whose group-wise average LCs are drawn in Fig.~\ref{simulation1}. Here merely the first two FPCs are capable of describing 82.965\% variation of the data set, which convincingly shows the effectiveness of FPCA for dimension reduction in LC study. Hierarchical clustering applied to these FPCs reveals the two inherent groups for which accuracy measure, the connectivity, attains its best possible value (Table \ref{t1}). In this case, clustering is achieved with 0\% misclassification. 
	
\begin{figure}
		\includegraphics[width=1\textwidth]{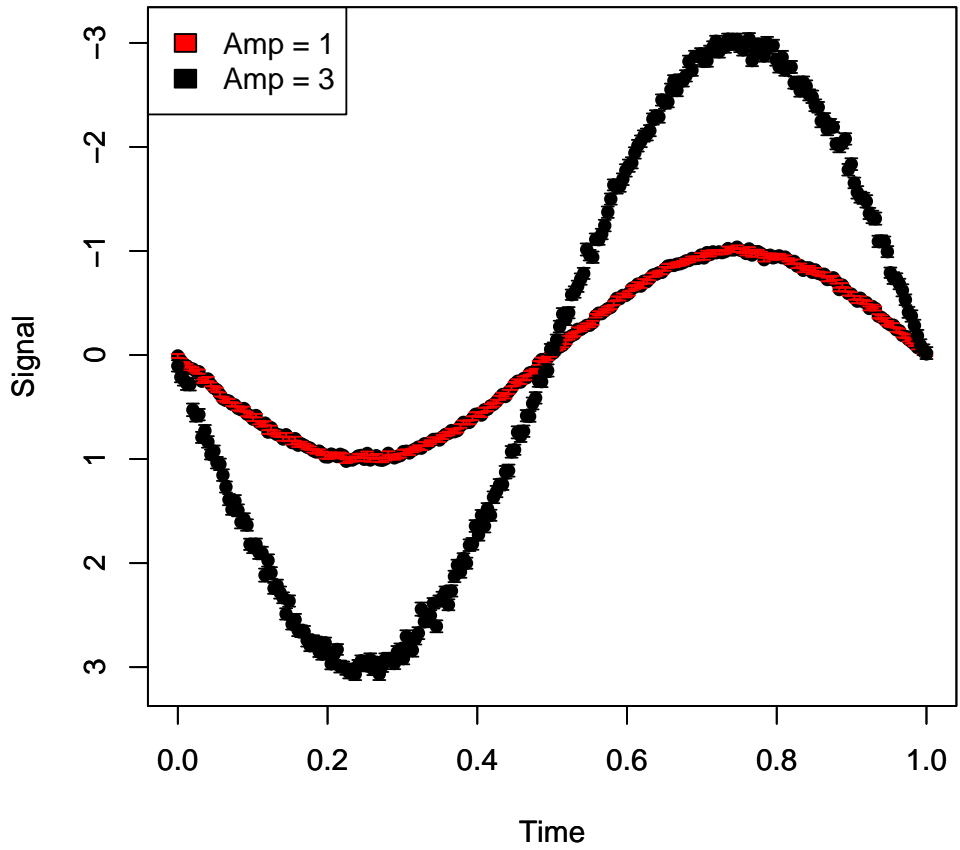}\\
		\caption{Simulated average light curves (with standard errors) of 500 members from each of the two groups generated by sine signals of amplitudes (amps) 1 and 3, with
			added noise and outliers}\label{simulation2}
	\end{figure}

	Next case generates LCs in the same manner using a complete cycle of the sine signal having two groups corresponding to amplitudes 1 and 3 (group-wise average LCs are given in Fig.~\ref{simulation2}). Here also clustering based on just the first two FPCs (explaining 66.278\% variation) exposes the two underlying groups of the data (Table \ref{t1}) with 0\% misclassification rate. 
	
\begin{figure}
	\includegraphics[width=1\textwidth]{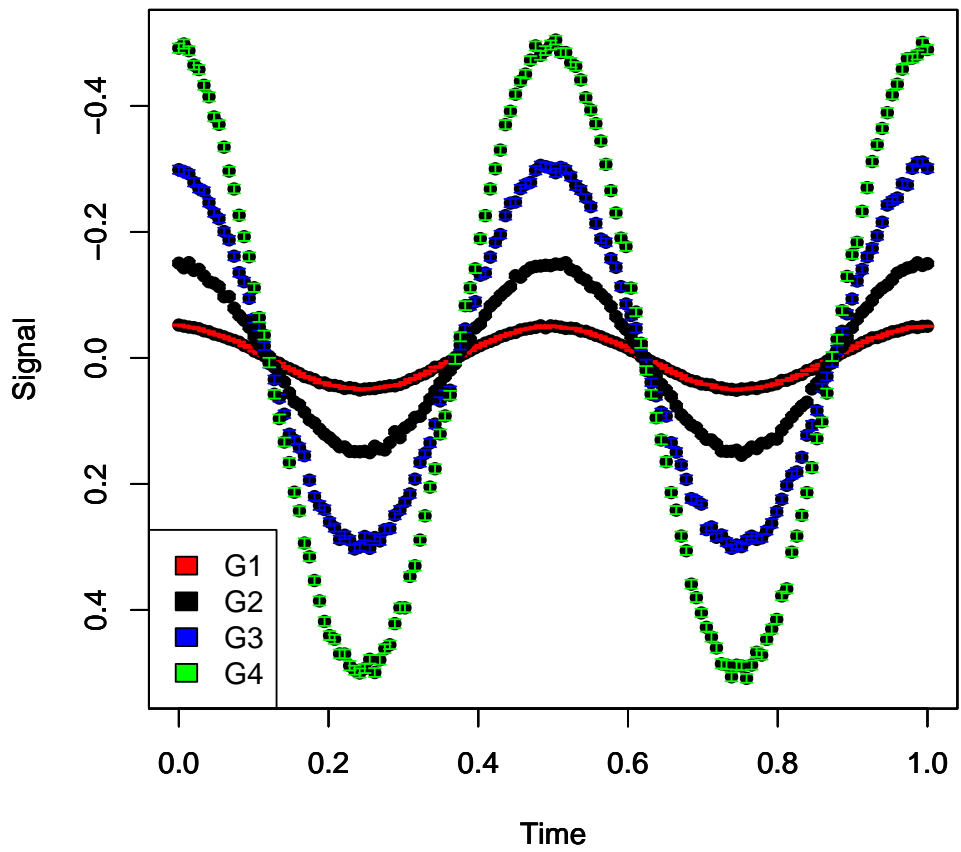}\\
	\caption{Simulated average light curves (with standard errors) for the members from four distinct groups (namely, G1--G4) of unequal sizes, generated by cosine signals with
		added noise and outliers.}\label{simulation3}
\end{figure}

	In the third simulation, an attempt has been made to produce LCs resembling more to the shape of the real-life LCs by creating synthetic curves at 150 equally spaced time points $t\in[0,1]$ using two consecutive cycles of the cosine signal added with 5\% outliers as follows (Fig.~\ref{simulation3}), where mentioned in order are the number of LCs, signal equation and the value of SNR added:\\
(i) 1000 LCs, signal = $-0.05 \cos(2\pi t)$, SNR = 1.5,\\
 (ii) 500 LCs, signal = $-0.15 \cos(2\pi t)$, SNR = 2,\\
  (iii) 450 LCs, signal = $-0.30 \cos(2\pi t)$, SNR = 2.5 and\\
   (iv) 850 LCs, signal = $-0.50 \cos(2\pi t)$, SNR = 3.\\
    Thus we have unequal number of LCs from four different groups with various amounts of noise. As the true classes of the LCs are known here, we use them to assess the performance of our method in revealing the original clusters in association with hierarchical clustering algorithm, carried out on the first two FPCs accounting for 69.794\% variation in the data, where for different number of clusters $k=2,3,4,5,6$, we respectively achieve 35.714, 35.714, 100.000, 87.071 and 75.607 \% correct classification. Therefore, our simulation study gives considerable evidence in establishing the proposed method as a potentially strong LC classifier which is very much useful to reduce the burden of big data sets significantly. Moreover, noise can be avoided from the data study by discarding the higher order FPCs. 
	\section{Results and discussion}
\begin{figure}
		\includegraphics[width=1\textwidth]{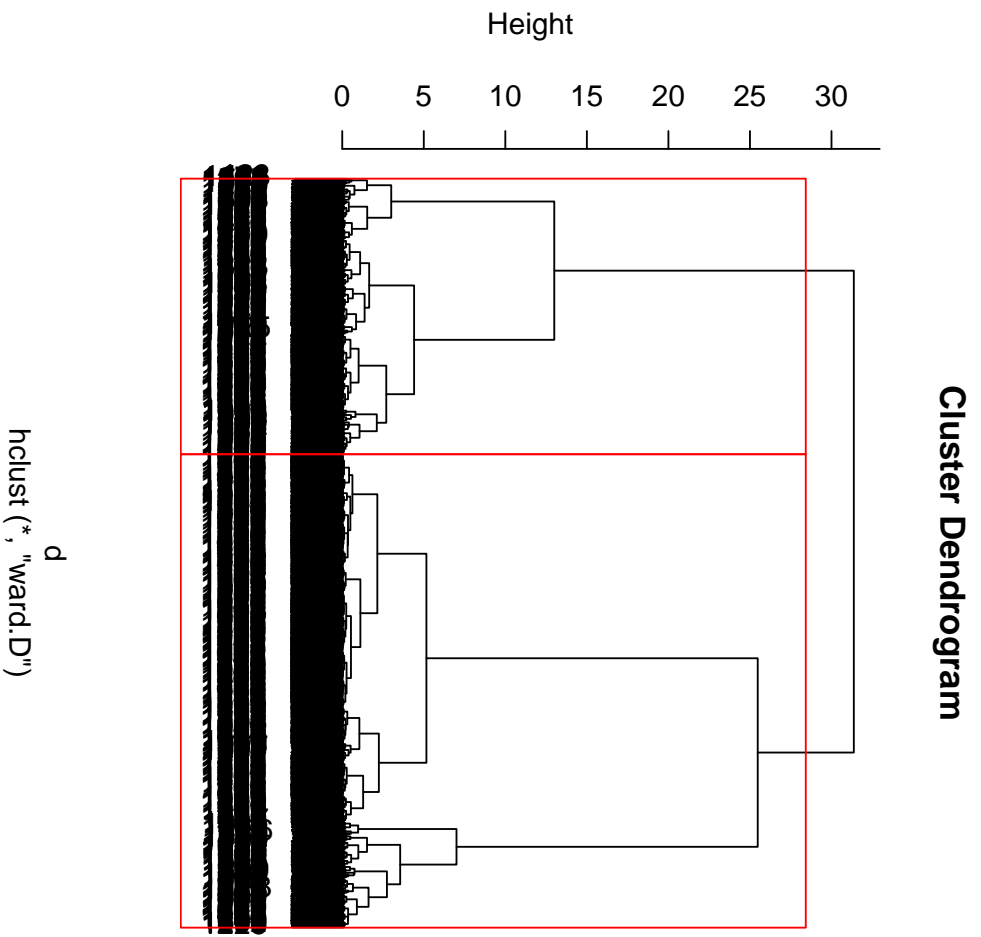}\\
		\caption{Dendrogram for Ward hierarchical clustering of the stars through FPCA shows two clusters k1 (right) and k2 (left).}\label{dendrogram}
	\end{figure}

\begin{figure}
		\includegraphics[width=1\textwidth]{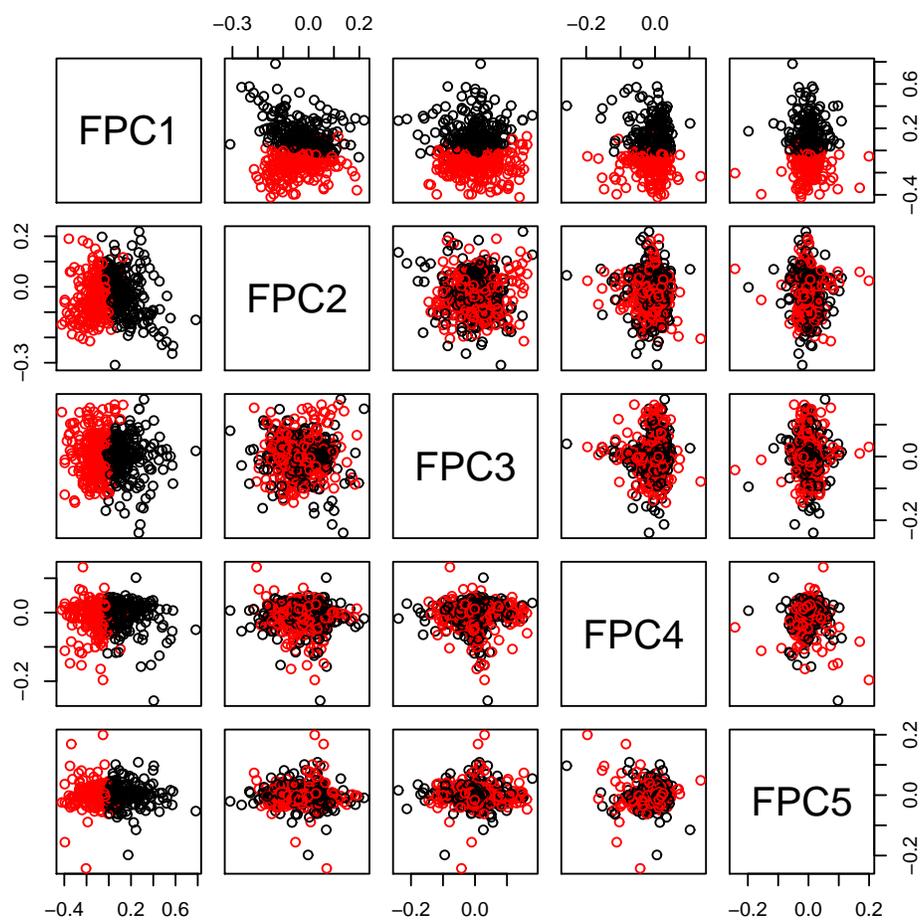}\\
		\caption{Matrix of scatter plots for each pair of the first five FPCs wherein components corresponding to the two clusters k1 and k2, obtained through FPCA, are shown in red and black colors respectively.}\label{scattermatrix}
	\end{figure}

\begin{figure}
		\centering
		\includegraphics[width=1\textwidth]{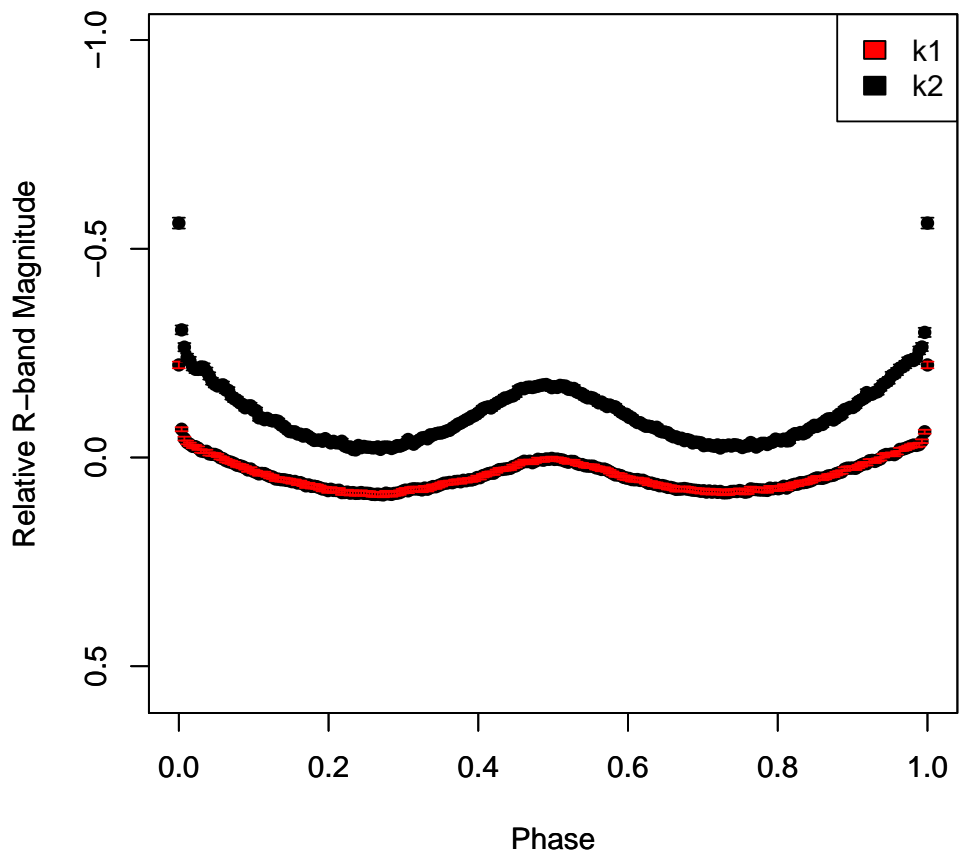}\\
		\caption{Template average light curves, with standard errors, of two clusters k1 and k2 obtained from FPCA.}\label{AvgLC}
	\end{figure}

	Hierarchical clustering based on the first five significant FPCs, explaining 83.699\% variation in the functional data, reveals two optimal clusters k1 and k2 (say) of sizes 833 and 485, respectively, in terms of the connectivity with the value 117.378 (Table \ref{t2}). It outperforms the clustering from Modak et al. (2020) which attained the connectivity equal to 149.69, as our present clustering produces a lower value for the clustering accuracy measure connectivity (see, Section 5.1); whereas the confusion matrix in Table~\ref{t3} studies an important comparison between the two clustering methods where both indicate two inherent clusters in the data set in terms of largely common stars falling in the two clusters k1 and k2 obtained through $k-$medoids method from Modak et al. (2020) and FPCA from this present work. Our cluster dendrogram obtained from hierarchical clustering is drawn in Fig.~\ref{dendrogram} which distinctly specifies the prevailing groups. We show a plot matrix, consisting of scatter plots for each pair of the first five FPCs, in Fig.~\ref{scattermatrix} whose first row and first column prominently visualize the two existing clusters. Cluster-wise subjective types from Miller et al. (2010) are given in Table~\ref{t4}, whereas Table~\ref{t5} and Fig.~\ref{AvgLC} respectively show the average properties and the average LCs for each cluster. Here two clusters, k1 LCs having lesser variation between the two minima of smaller depths (Fig.~\ref{AvgLC}) and higher average time period (Table~\ref{t5}) than those in k2, are consistent with the resulted two groups in Modak et al. (2020) obtained through $k$-medoids clustering with CID (see, Figs~5--6 and Table 4 in Modak et al. 2020). Both clusterings indicate two groups of Es irrespective of the subjective classification by Miller et al. (2010) (see, Tables~3 and \ref{t4} respectively from Modak et al. 2020 and the present paper). It is to be noted that the template medoid LCs (Fig~6 in Modak et al. 2020) and the representative observed LCs (Figs~7--8 in Modak et al. 2020) robustly fall in our present clusters k1 and k2. Moreover, in Modak et al. (2020), clusters g1 and g2 containing the distinct LCs obtained from the fuzzy k-means clustering (Bezdek 1981), which have more separated average LCs for two groups (Fig~9 in Modak et al. 2020) than the overlapping LCs found in $k$-medoids clustering with CID (Figs~5--6 in Modak et al. 2020), pretty much resemble the present results achieved from FPCA (Fig.~\ref{AvgLC}). It clearly shows the superiority of our current proposed method, in finding distinct clusters of Es based on their LCs, over our previously implemented method in Modak et al. (2020). 
	
\begin{figure}
		\includegraphics[width=1\textwidth]{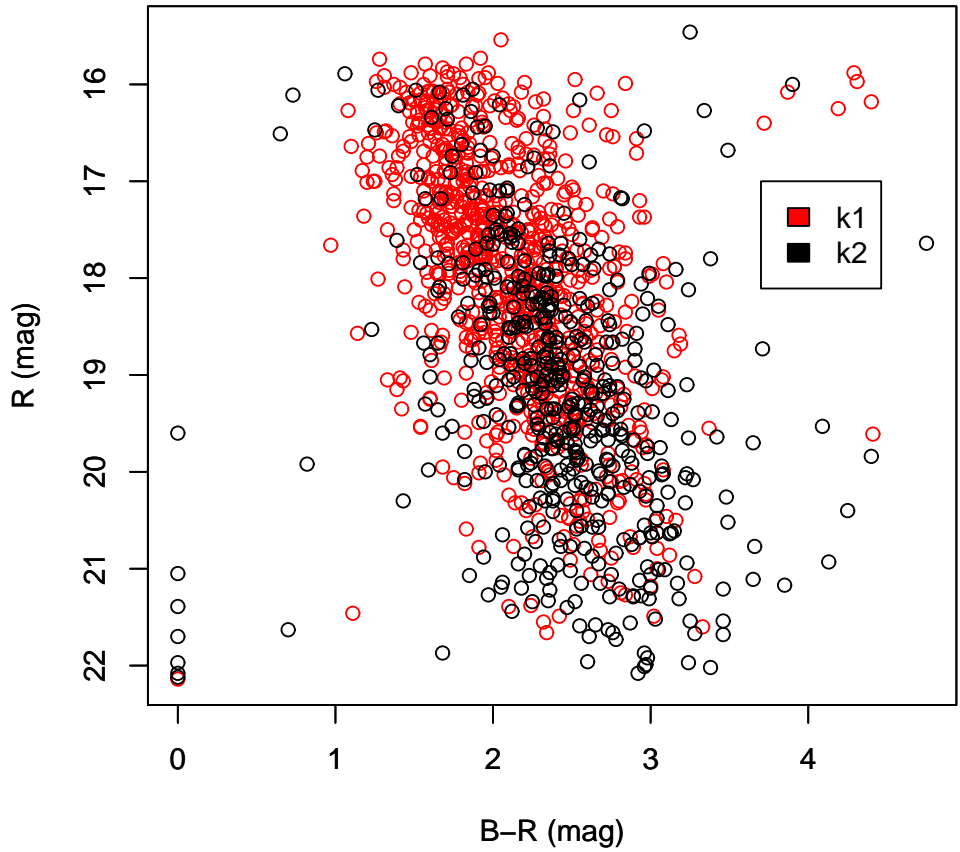}\\
		\caption{Color-magnitude (B-R versus R) diagram of the stars clustered in two groups k1 and k2 through FPCA.}\label{ColorDiagram}
	\end{figure}

\begin{figure}
		\includegraphics[width=1\textwidth]{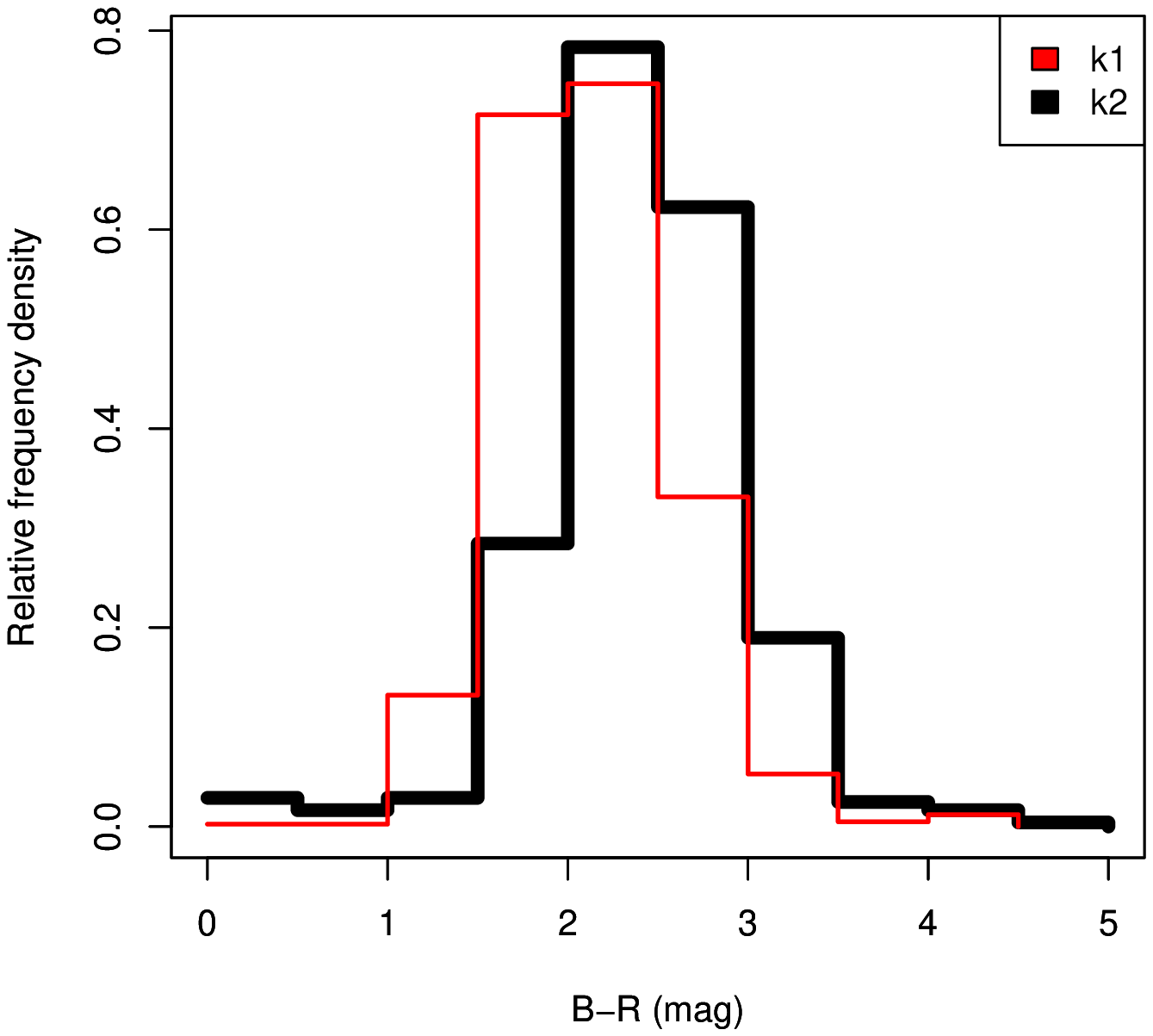}\\
		\caption{Histograms of B-R color index for the stars clustered in two groups k1 and k2 through FPCA.}\label{Hist_B-R}
	\end{figure}

\begin{figure}
		\includegraphics[width=1\textwidth]{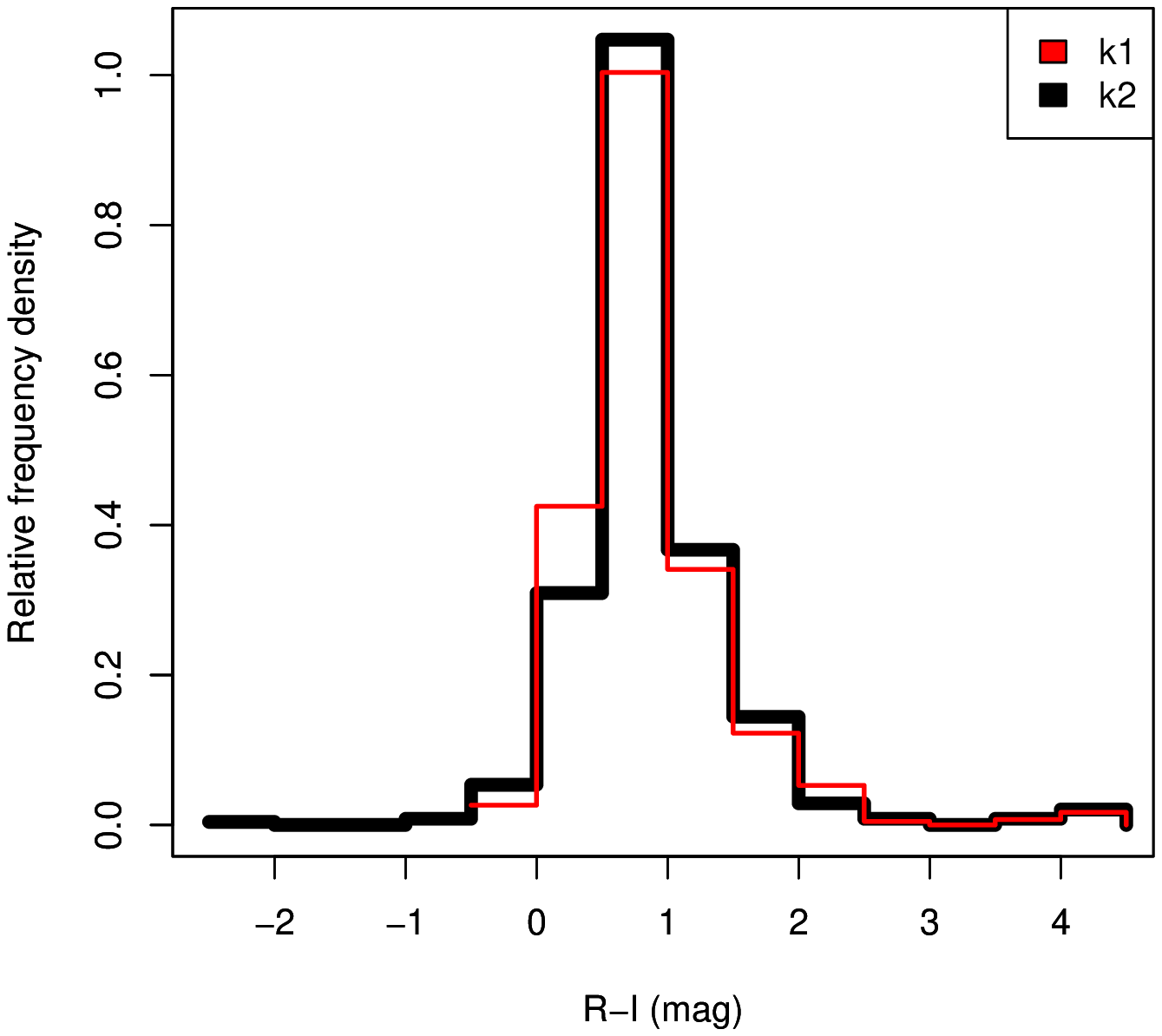}\\
		\caption{Histograms of R-I color index for the stars clustered in two groups k1 and k2 through FPCA.}\label{Hist_R-I}
	\end{figure}

\begin{figure}
	\includegraphics[width=1\textwidth]{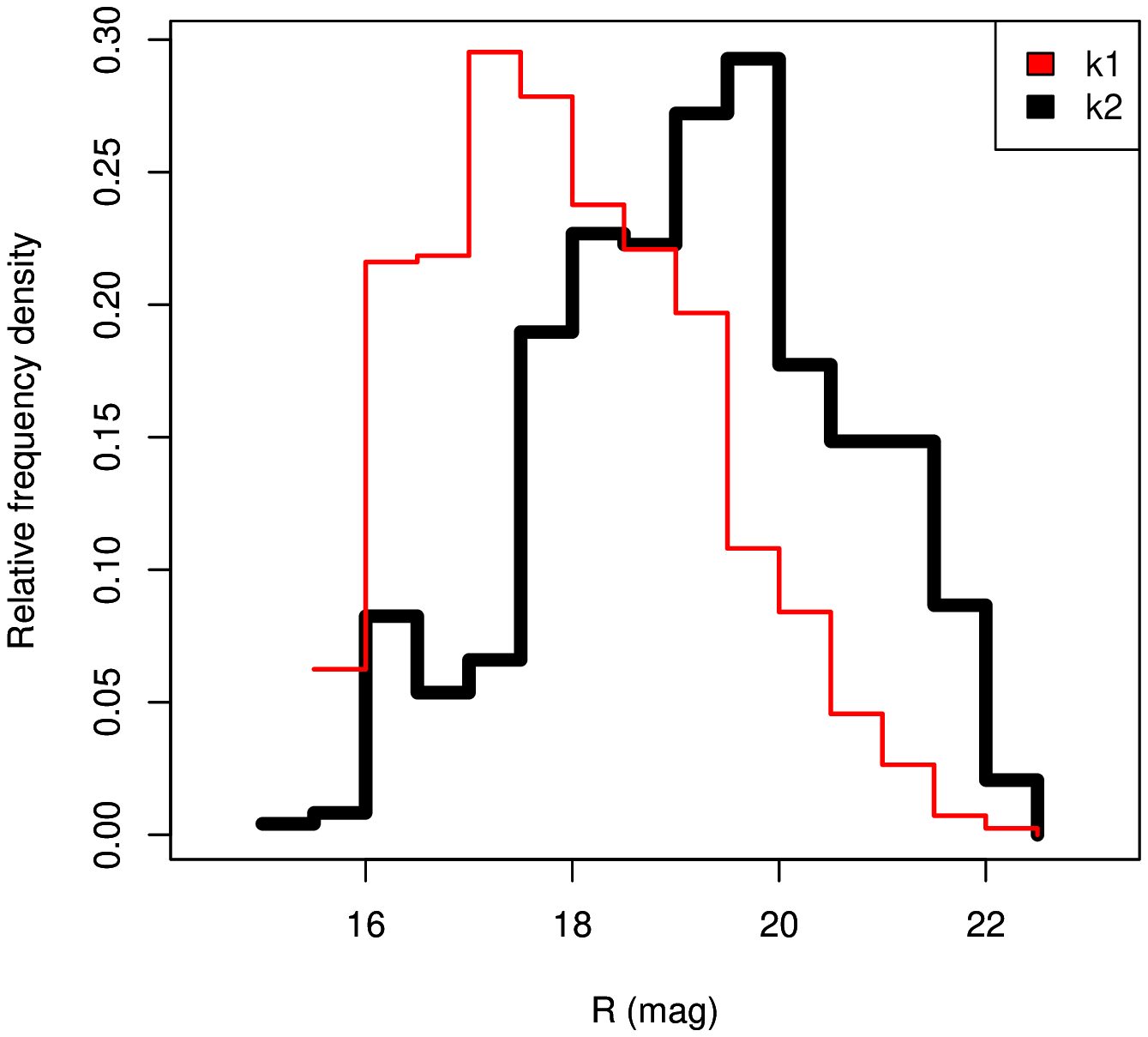}\\
	\caption{Histograms of R magnitude for the stars clustered in two groups k1 and k2 through FPCA.}\label{Hist_R}
\end{figure}

\begin{figure}
	\includegraphics[width=1\textwidth]{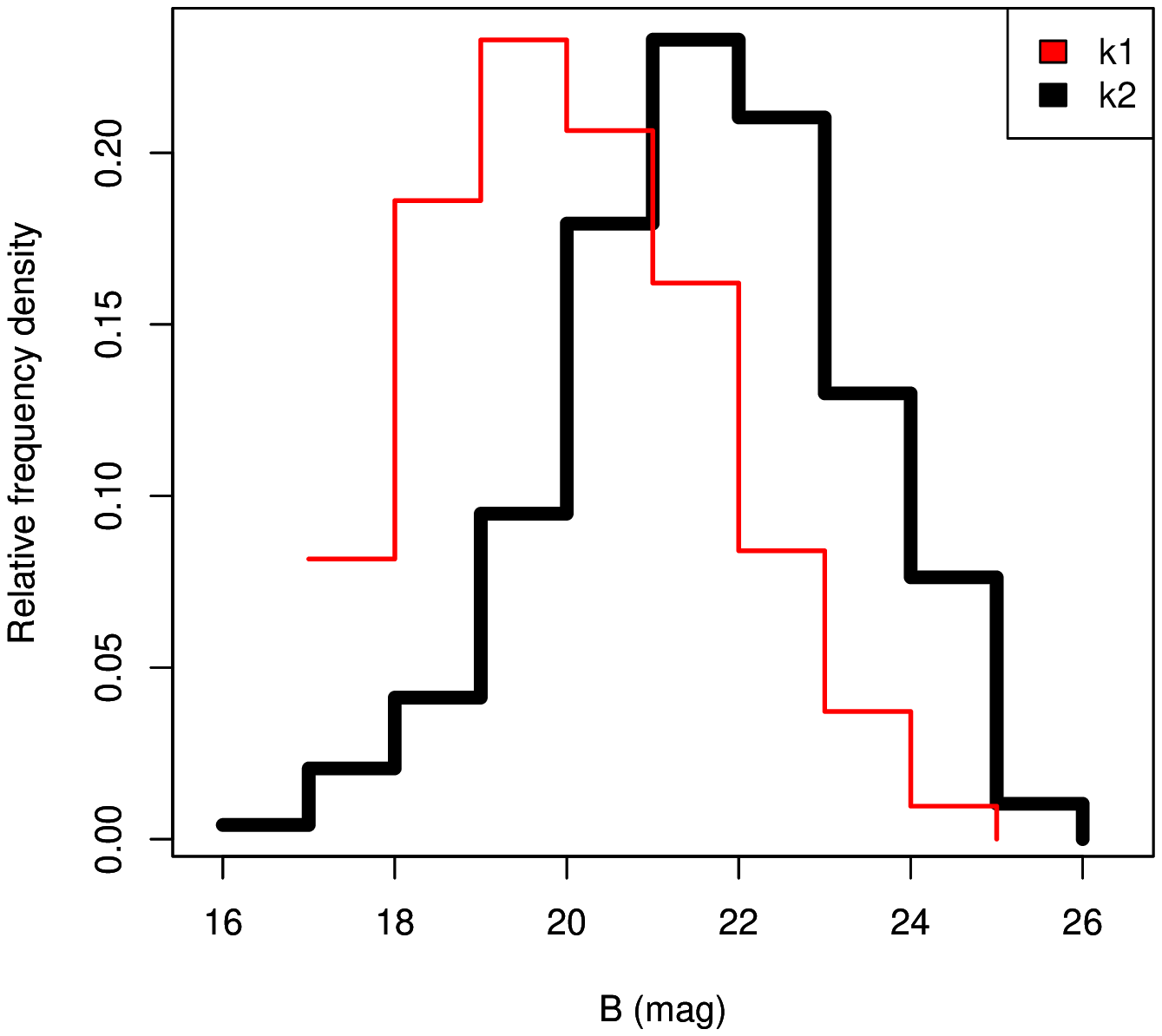}\\
	\caption{Histograms of B magnitude for the stars clustered in two groups k1 and k2 through FPCA.}\label{Hist_B}
\end{figure}

\begin{figure}
	\includegraphics[width=1\textwidth]{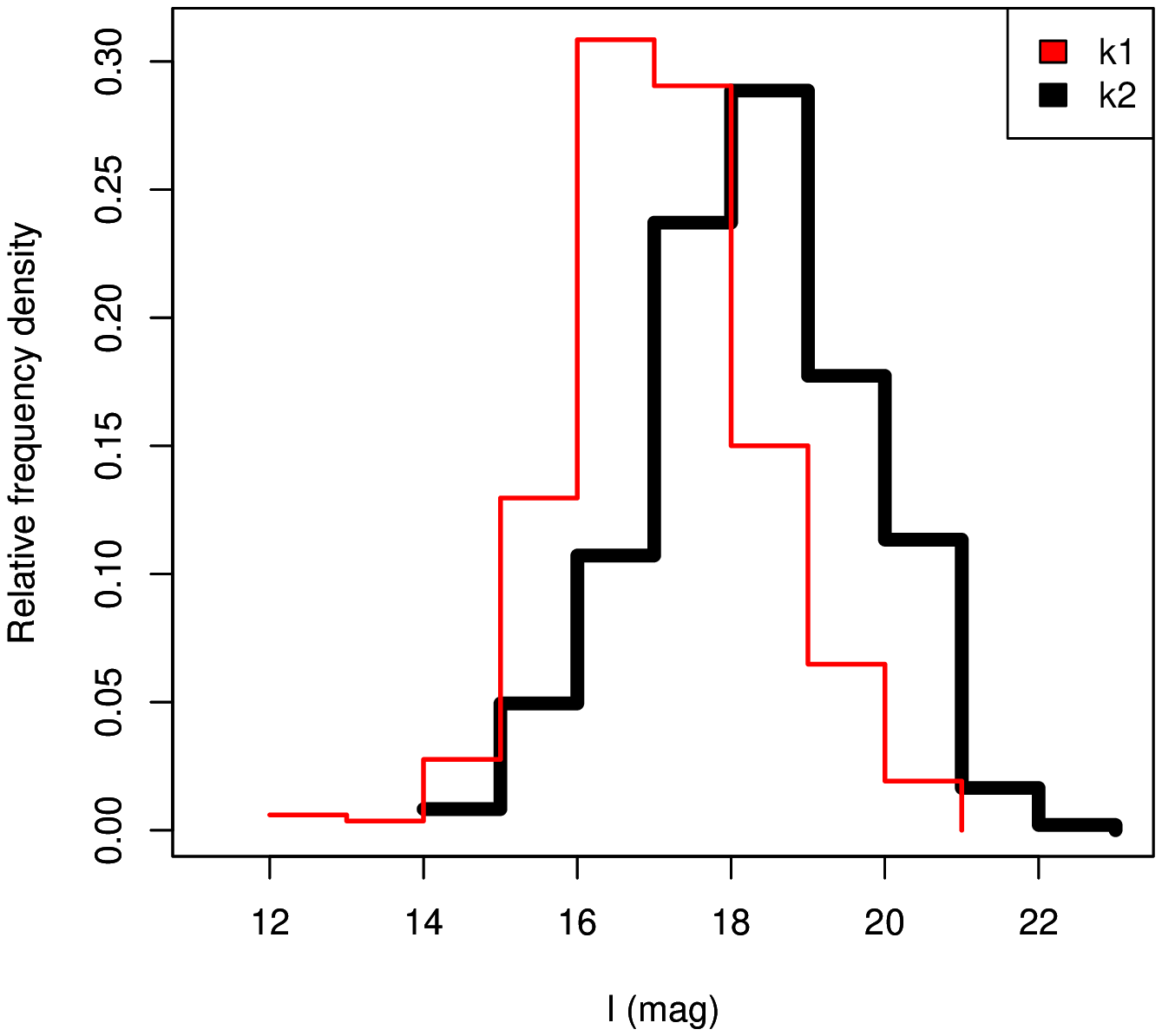}\\
	\caption{Histograms of I magnitude for the stars clustered in two groups k1 and k2 through FPCA.}\label{Hist_I}
\end{figure}

	The LCs belonging to the first cluster k1 exhibit less variation between the two minima in comparison to those in the second cluster k2 (Fig.~\ref{AvgLC}), whereas k1 LCs possess higher period on an average than k2 curves (Table~\ref{t5}). These indicate k1 systems to be formed of a more or less detached or semidetached. Moreover, the depths of the two minima of the LCs are lower for k1 than for k2, which states a k1 binary system is having stars of unequal masses with a less massive secondary. On the other hand, the two stars of a k2 system are almost comparable in their masses. Color-magnitude diagram of the stars (Fig~\ref{ColorDiagram}) and group-wise histograms for different color indices (Figs~\ref{Hist_B-R} and \ref{Hist_R-I}) disclose two overlapping clusters with most of the k1 stars having larger R magnitudes (i.e. smaller R values, where a lower numerical value means higher R magnitude) while comparing to k2 stars; whereas Table 5 shows slightly higher average B--I and R--I magnitudes for the first cluster. Also, Table~\ref{t5} and histograms for R, B, I magnitudes (Figs.~\ref{Hist_R}--\ref{Hist_I}) expose that k1 stars are possessing the mentioned properties with larger averages and higher modes in their magnitudes in contrast to k2 stars. All these facts suggest connection of k1 systems (bluer) with higher temperature to early spectral type and k2 systems (redder) to late spectral type. 
	\section{Conclusion}
	This paper clusters 1318 variable stars based on their observed LCs in a new approach where discrete time series are treated as continuous curves over time in the form of functional data. Then FPCA extracts relevant information from the functional data curves, where we considerably reduce the dimension and discard the noise simultaneously in terms of the first five significant FPCs. A simple classical statistical method like hierarchical clustering is good enough to expose the true clusters in terms of the important FPCs. It confirms the existing two groups of Es in a more prominent way than the clustering of Modak et al. (2020). The conspicuously convincing results of simulation study establish our approach as an effective LC-based classifier efficient enough to handle big data sets.
	\section*{Acknowledgements}
	The authors would like to thank the editors for encouraging the present work on Astrostatistics and one anonymous reviewer for its intriguing inquiries which helped the authors to present the results in a more convincing way.
	\clearpage
	\section*{Data Availability} All data analyzed and generated during this study are referenced in this published article.
	\section*{Conflicts of Interest}
	No potential conflict of interest was reported by the authors.
	\clearpage
	\begin{table}
		\caption{Computed values of the connectivity for different number of clusters ($k$) from FPCA for the first two simulated data sets considered}
		\label{t1}
		\begin{tabular}{lll}
			\hline\noalign{\smallskip}
			$k$ & Connectivity & Connectivity \\
			& data set 1&data set 2\\ \noalign{\smallskip}\hline\noalign{\smallskip}
			2 &  0 &0\\
			3 & 23.354 &29.488\\
			4 & 55.458 & 54.120\\
			5 & 73.258 &62.814\\
			6 & 78.393 &73.175\\
			\noalign{\smallskip}\hline
		\end{tabular}
	\end{table}
	\begin{table}
		\caption{Computed values of the connectivity for different number of clusters ($k$) from FPCA for the LC data set under study}
		\label{t2}
		\begin{tabular}{ll}
			\hline\noalign{\smallskip}
			$k$ & Connectivity \\ \noalign{\smallskip}\hline\noalign{\smallskip}
			2 & 117.378 \\
			3 & 168.298 \\
			4 & 210.842 \\
			5 & 225.409 \\
			6 & 309.082 \\
			\noalign{\smallskip}\hline
		\end{tabular}
	\end{table}

\begin{table}
	\caption{Confusion matrix comparing the membership of the stars in two clusters k1 and k2 resulted in two different methods, namely the present one using FPCA and $k-$medoids clustering from Modak et al. (2020)}
	\label{t3}
	\begin{tabular}{|l|l|l|l|}
		\hline
		   Cluster  &k1 (FPCA) & k2 (FPCA)&Total\\ \hline
	k1 ($k-$medoids)&633       &205       &838\\
	k2 ($k-$medoids)&200       &280       &480\\\hline
	Total           &833       &485       &1318\\ 
	\hline	
	\end{tabular}
\end{table}	
	\begin{table}
	\caption{Membership of subjective types according to Miller et al. (2010) in two clusters k1 and k2 obtained from FPCA}
	\label{t4}
	\begin{tabular}{lll}
		\hline\noalign{\smallskip}
		Type &k1 &k2\\
		\noalign{\smallskip}\hline\noalign{\smallskip}
		EA&36&51\\
		EB& 78&  30\\
		EW&264& 230\\
		PUL&94&24\\
		EA:&16&30\\
		EB:&155&60\\
		EW:&7&14\\
		PUL:&142&26\\
		CV:&1&0\\
		EA/EB&5&5\\
		EW/EA&2& 4\\
		EW/EB&5&7\\
		EB/PUL&10&2\\
		DCEP/PUL&9&0\\
		CV/PUL&9&2\\[.5 ex]
		\hline
		\noalign{\vskip .05 in}
		Total&833&485\\	
		\noalign{\smallskip}\hline
		Note: An uncertain type is followed by a
		colon and\\ an ambiguous type is given with a slash.
	\end{tabular}
\end{table}
\clearpage
	\begin{table}
		\tiny
		\caption{Average values (with standard errors) of the variables for stars in two clusters k1 and k2 obtained from FPCA}
		\label{t5}
		\begin{tabular}{llllllll}
			\hline\noalign{\smallskip}
			Name of  & No. of  &P & R                & B                & I                & B-I               & R-I \\ [0.5ex]
			cluster  & stars &(day) &(mag) &(mag) &(mag) &(mag) &(mag)\\
			\noalign{\smallskip}\hline\noalign{\smallskip}
			k1     &833
			&2.758$\pm$0.130       & 18.017$\pm$0.045 & 20.144$\pm$0.054 & 17.153$\pm$0.044 & 2.991$\pm$0.028 & 0.864$\pm$0.020 \\
			k2     & 485
			&1.514$\pm$0.117       & 19.214$\pm$0.064 & 21.635$\pm$0.076 & 18.323$\pm$0.062 & 3.312$\pm$0.039 & 0.891$\pm$0.028
			\\
			\noalign{\smallskip}\hline
		\end{tabular}
	\end{table}	
	\clearpage
	

\end{document}